\documentclass[%
 reprint,
superscriptaddress,
 aps,
floatfix,
]{revtex4-2}

\usepackage{graphicx}
\usepackage{xcolor}
\usepackage{dcolumn}
\usepackage{bm}
\usepackage{comment}
\usepackage{wrapfig}
\usepackage{natbib}
\usepackage{multirow}
\usepackage{makecell}
\usepackage{float}
\usepackage{amsmath}
 
\usepackage{units}
\usepackage{color}
\usepackage{url}

\usepackage[colorlinks]{hyperref}
\hypersetup{%
	plainpages=true,
	breaklinks=true,
	hypertexnames=false,
	pageanchor=true,
	colorlinks=true,
	linkcolor={blue},
	citecolor={red},
	urlcolor={blue},
	anchorcolor={black}
}

\newcommand{\ket}[1]{|#1\rangle}

\begin{document}

\preprint{APS/123-QED}

\title{Quantum refrigeration powered by noise in a superconducting circuit
} 

\author{Simon Sundelin}
 \email{simsunde@chalmers.se}
\author{Mohammed Ali Aamir}
\author{Vyom Manish Kulkarni}
\author{Claudia Castillo-Moreno}
\author{Simone Gasparinetti}
\email{simoneg@chalmers.se}
 \homepage{https://202q-lab.se}
\affiliation{%
Department of Microtechnology and Nanoscience, Chalmers University of Technology, 412 96 Gothenburg, Sweden
}%

\date{\today}

\begin{abstract}
While dephasing noise frequently presents obstacles for quantum devices, it can become an asset in the context of a Brownian-type quantum refrigerator.
Here we demonstrate a novel quantum thermal machine that leverages noise-assisted quantum transport to fuel a cooling engine in steady state. The device exploits symmetry-selective couplings between a superconducting artificial molecule and two microwave waveguides. These waveguides act as thermal reservoirs of different temperatures, which we regulate by employing synthesized thermal fields. We inject dephasing noise through a third channel that is longitudinally coupled to an artificial atom of the molecule.
By varying the relative temperatures of the reservoirs, and measuring heat currents with a resolution below 1 aW, we demonstrate that the device can be operated as a quantum heat engine, thermal accelerator, and refrigerator. Our findings open new avenues for investigating quantum thermodynamics using superconducting quantum machines coupled to thermal microwave waveguides. 

\end{abstract}

\maketitle


As quantum technologies advance, so does the imperative to understand energy flows at the quantum level. The extension of thermodynamics to single quantum systems has uncovered fundamental insights into nanoscale out-of-equilibrium systems and the second law of thermodynamics \cite{lieb1999a, janzing2000a, egloff2015a, horodecki2013a, brandao2015a, yungerhalpern2016b, halpern2018a, lostaglio2017a, guryanova2016a, sparaciari2017a}. This progress has also unveiled new avenues for boosting the efficiency of batteries \cite{binder2015a} and for the optimization of quantum heat engines \cite{kalaee2021, hammam2022a, lostaglio2020a, myers2020a, kim2011d}. Of particular interest is the category of thermal machines that operate autonomously by harnessing heat flows to perform beneficial tasks. In the quantum domain, these machines provide an ideal setting for measuring the thermodynamic expenditure associated with operations such as timekeeping \cite{erker2017a}, entanglement generation \cite{tavakoli2018a, brask2015b} and refrigeration \cite{linden2010a, levy2012a, chen2012h, venturelli2013a, correa2014a, silva2015a, hofer2016c, mu2017a, maslennikov2019a, du2018a, holubec2019b, mitchison2019a, bhandari2021a, kloc2021a, almasri2022a}.

In many quantum technologies and experiments, cooling serves as an essential preliminary step. Consequently, quantum absorption refrigerators represent a notable subset of thermal machines, distinguished by their capacity to produce valuable outputs solely through the utilization of heat as a resource. The primary goal of such a refrigerator is to transfer heat from a cold bath of temperature $T_c$ to a hot bath of temperature $T_h > T_c$ by using heat from a work reservoir at a temperature $T_w > T_h$ \cite{mitchison2019b}. The earliest and arguably simplest model of a quantum absorption refrigerator is the three-level system \cite{scovil1959a, geusic1967, palao2001a}, with the more intricate three-body refrigerator recently realised using trapped ions \cite{maslennikov2019a} and superconducting qubits \cite{aamir2023b}.

Compared to how absorption refrigerators utilize heat from a work reservoir to operate autonomously, a Brownian refrigerator can do so by harnessing thermal fluctuations \cite{vandenbroeck2006a, pekola2007a}. Analogous to the operation of Brownian motors \cite{parrondo2002, reimann2002, astumian2002}, these devices facilitate unidirectional heat transfer in response to random noise. This distinction positions Brownian refrigerators as a noteworthy category of thermal machines, highlighting their unique operational mechanism. However, the development of a cooling system based on this principle has proven to be elusive because of the difficulty in mediating energy flows through thermal fluctuations. In the quantum domain, this form of energy transport is akin to noise-assisted excitation transport observed in quantum networks \cite{creatore2013a, rebentrost2009a, chin2012a, rey2013a}. Such transport is believed to be a governing mechanism in the process of photosynthesis and its viability has been investigated in a three-site network \cite{potocnik2018c}. Beyond leveraging this mechanism to function against a temperature gradient, realizing a Brownian refrigerator necessitates precise measurements of minute heat currents over a varying temperature gradients, a technically challenging task.
\begin{figure*}
    \includegraphics[width=17.5cm]{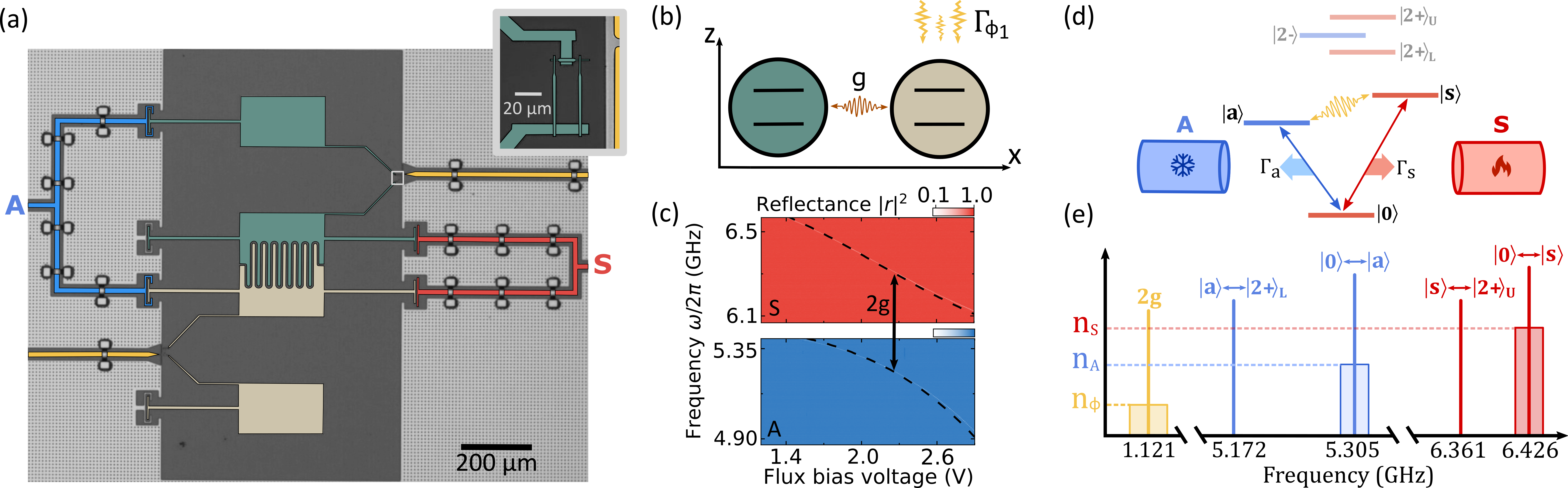}
    \caption{\label{fig1} Device architecture and energy level diagram. (a) False-color micrograph of the device comprised of two frequency-tunable transmon, colored in green and beige, coupled to microwave waveguides labeled S (red) and A (blue). Flux lines coupling longitudinally to the system are colored in yellow. Inset shows a superconducting quantum interference device (SQUID) of one of the transmons.  (b) The transmons represent two coupled qubits where each can be dephased through a longitudinally coupled channel. (c) Reflectance, $|r|^2$, through symmetric (S) and antisymmetric (A) waveguide as a function of applied flux voltage bias. The coupling rate, $g$, is obtained at the point where $\Delta = \omega_s-\omega_a = 2g$ is minimized. (d) Energy-level diagram showing states of even (red) and odd (blue) symmetry up to second-excitation manifold of the artificial molecule realized with the transmons. Symmetry-preserving (red) and symmetry-inverting (blue) transitions predominantly couple to waveguide S and A, respectively. For clarity, transitions to higher excited states have been omitted. The effect of the bare qubit dephasing in the collective state picture is to mix states $|a\rangle$ and $|s\rangle$. (e) Experimentally observed transition frequencies of the molecule in the first excitation-manifold along with the next closest transition. The red, blue and yellow shaded boxes represents the spectral density of the injected thermal radiation corresponding to thermal populations of $n_S$, $n_A$ and $n_{\Phi}$ in waveguide S, A and one of the flux lines.}
\end{figure*}

In this Letter, we present the experimental realization of an autonomous Brownian-type quantum refrigerator with a superconducting circuit. The refrigerator is driven by thermal fluctuations via the process of noise-assisted excitation transport. We directly observe heat currents on the order of 1 aW by performing simultaneous power spectral density (PSD) measurements through two waveguides acting as thermal baths.
We show that the power transfer between the baths is dictated by the amount of thermal noise we inject. In addition, we independently regulate the temperature of the baths by employing synthesized thermal fields. Depending on the relative temperature of the baths,
we demonstrate how our device can operate as an autonomous quantum refrigerator, heat valve or thermal accelerator.

Our device is comprised of two nominally identical, flux-tunable transmon qubits \cite{koch2007b}, forming an artificial molecule. Each transmon consists of two superconducting islands which form the capacitor and are shunted by a superconducting quantum interference device (SQUID), to which we couple a flux line [Fig.~\ref{fig1}(a)].
The Hamiltonian governing the system can be expressed as \\
$H=$ $\sum_{i=1,2}\omega_i(\Phi_i) \sigma_i^{+} \sigma_i^{-}+g\left(\sigma_1^{+}\sigma_2^{-}+\sigma_2^{+}\sigma_1^{-}\right)$. Here $\omega_i(\Phi_i)$ are the bare, flux-tunable mode frequencies, $\sigma_i^{+}$ and $\sigma_i^{-}$ are the creation and annihilation operators of qubit $i = {1,2}$ respectively, and $g$ is the coupling rate between them \cite{koch2007b}. When the bare mode frequencies are equal, the states $|10\rangle$ and $|01\rangle$ are resonant and in the molecule's single-excitation manifold they form the collective states $|s\rangle=(|10\rangle+|01\rangle) / \sqrt{2}$ and $|a\rangle=(|10\rangle-|01\rangle) / \sqrt{2}$. These are symmetric and antisymmetric, respectively, and their frequencies are split by 2$g$ [Fig \ref{fig1}. (d)].Two microwave waveguides, denoted by S and A, are capacitively coupled to multiple points of the circuit to predominantly facilitate symmetry-preserving (waveguide S) and symmetry-inverting transitions of the molecule (waveguide A), with respect to a permutation of the two artificial atoms~\cite{aamir2022a}.
Hence, in the first excitation-manifold, the $|0\rangle \leftrightarrow|a\rangle$ transition largely couples to waveguide A, while the $|0\rangle \leftrightarrow|s\rangle$ transition couples to waveguide S. Simultaneously, flux lines longitudinally couple to the bare qubits, thereby creating the possibility to induce controlled dephasing through the application of noise. 
 
To increase the sensitivity of the molecular modes to applied flux noise, we tune the transition frequency of qubit 1 away from its zero flux point (``sweet spot''). Next, we tune the magnetic flux in the second flux line until $\Delta = \omega_s-\omega_a$ is minimized [Fig.~\ref{fig1}(c)]. This point correspond to the full hybridization of the qubits where $\omega_1=\omega_2$, ensuring maximal isolation between the symmetric (antisymmetric) mode with the unintended antisymmetric (symmetric) waveguide. The frequencies of the resulting modes are determined to be $\omega_s /2\pi = 6.426$~GHz and $\omega_a /2\pi = 5.305$~GHz, indicating a coupling rate between the transmons of $g/2\pi = 560$~MHz [Fig \ref{fig1} (c)]. 

\begin{figure}
    \includegraphics[width=8.4cm]{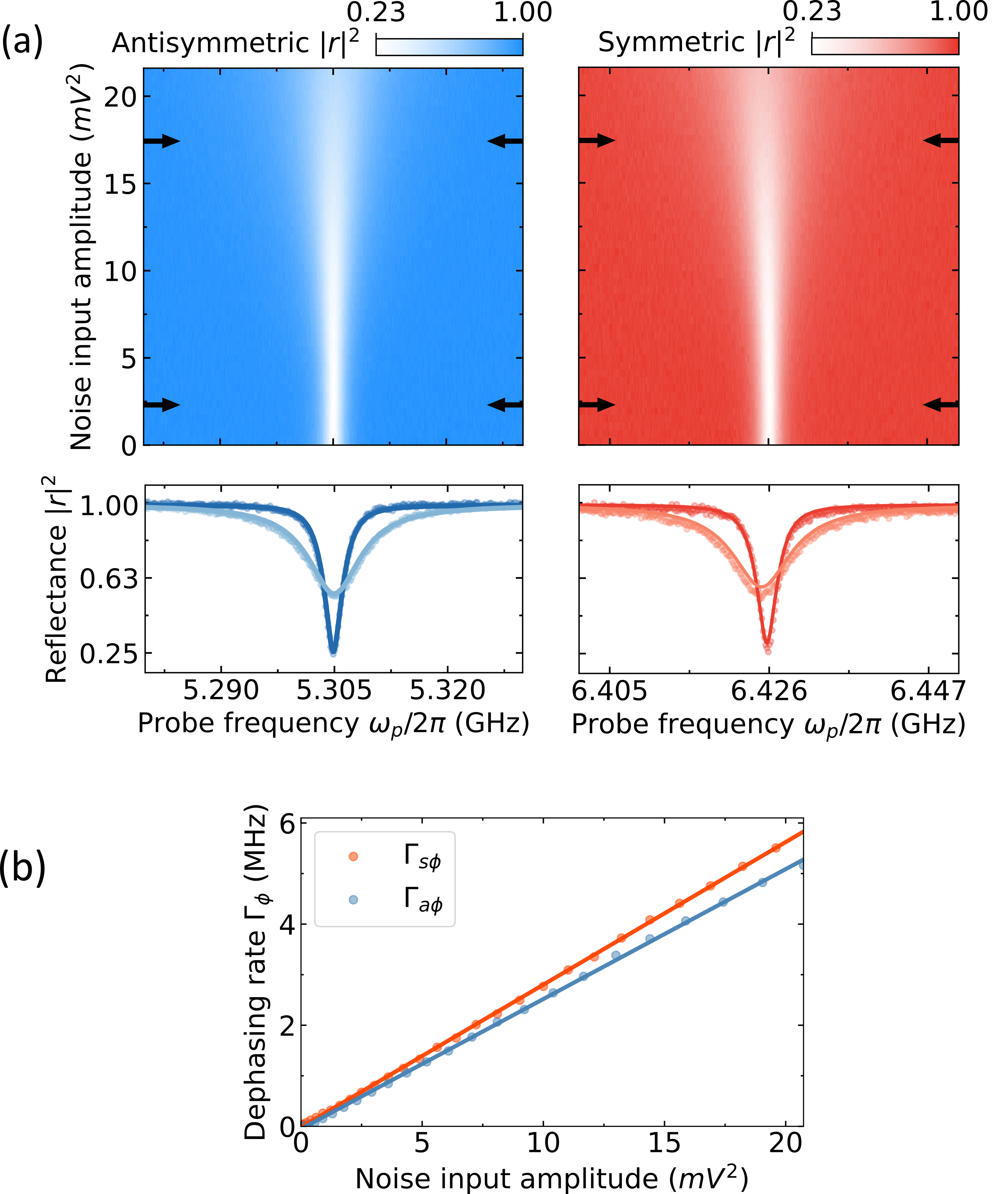}
    \caption{\label{fig2} Dephasing rate characterization through both waveguides. (a) 2D plot of power reflectance, $|r|^2$, as a function of probe frequency $\omega_p$ and injected noise power. Line cuts of reflectance at input noise powers indicated by the arrows in. Solid lines are fits based on our theoretical model (see Sec.~II of Supplementary Information). (b) Dephasing rate of both symmetric and antisymmetric mode dependent on input noise power.}
\end{figure}

To determine the coupling rates of the modes into the waveguides, we measure the reflection coefficient, $r$, from each waveguide as a function of frequency and power, and globally fit a model based on Lindblad master equation and input-output theory to the data \cite{aamir2022a} (see Sec.~II of Supplementary material).

We determine the radiative coupling rates of the symmetric (antisymmetric) mode to the respective waveguide to be $\Gamma_s/2\pi = 2.87$ MHz ($\Gamma_a/2\pi = 2.83$ MHz). At the same time, we find additional coupling rates of $\Gamma_s^{\prime}/2\pi = 98$ kHz ($\Gamma_a^{\prime}/2\pi = 97$ kHz) to other channels, which we ascribe to unintended coupling of each mode to the opposite waveguide. Notably, the transition $|0\rangle \leftrightarrow|s\rangle$ ($|0\rangle \leftrightarrow|a\rangle$) exhibits overcoupling to waveguide S (A) with a selectivity ratio of $\Gamma_s/\Gamma_s^{\prime} = 29$ ($\Gamma_a / \Gamma_a^{\prime} = 29$).

We first study the impact of dephasing noise on the $|0\rangle \leftrightarrow|s\rangle$ and $|0\rangle \leftrightarrow|a\rangle$ transitions. To do so, we apply filtered white noise to the flux line connected to qubit 1, with a flat spectral profile $S_{\Phi}(\omega)$ characterized by a 50 MHz bandwidth centered at $2g$ and tunable amplitude. For increasing noise power, the reflectance exhibits broadening of the linewidth in both the antisymmetric and symmetric mode [Fig. \ref{fig2} (a)]. Employing the same theoretical framework used for determining the coupling rates, the dephasing rate for varying noise power can be determined from a global fit with only the frequency and $\Gamma_{\phi}$ as variable parameters [Fig. \ref{fig2} (b)]. The observed linewidth broadening primarily stems from the frequency components of the applied noise spectrum $S_{\Phi}(\omega)$ that bridges the energy gap between state $|s\rangle$ and $|a\rangle$. We ascertain this from the observation that substantial broadening is only noticeable when the spectral profile of the flux noise overlaps in frequency with $2g$. Furthermore, upon varying the bandwidth of the noise for a fixed amplitude, the linewidth saturates when the noise bandwidth exceeds the mode linewidth $\Gamma_{\{s,a\}}$ (see Sec.~III of Supplementary Material).

\begin{figure}
    \includegraphics[width=8.4cm]{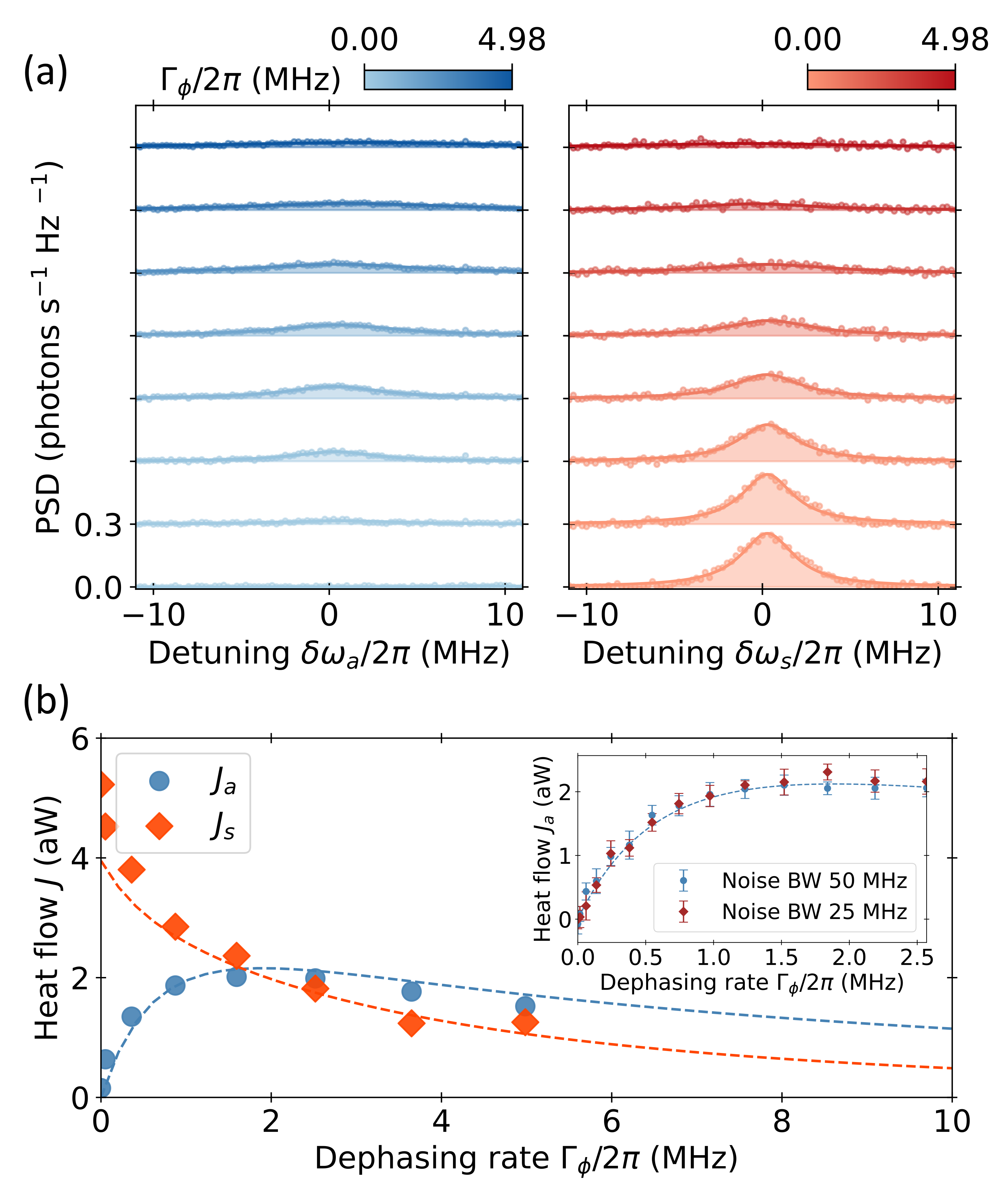}
    \caption{\label{fig3} Measured power transfer between waveguides for coherent excitation of the symmetric mode at the Rabi frequency $\Omega_s/2\pi = 1.47$ MHz. (a) Power spectral density (PSD) extracted simultaneously from both the antisymmetric (blue) and symmetric (red) waveguide for different dephasing rates.  Contributions from elastic scattering have been subtracted from the data. (b) Heat flow obtained from the integrated PSD from both waveguides as a function of dephasing. The inset highlights the power into the antisymmetric waveguide under two distinct noise bandwidths. Dashed lines represent the result of master equation simulations with independently extracted parameters (see Sec.~VI of Supplementary Material).}
    \label{figure1}
\end{figure}

We demonstrate dephasing-mediated energy transfer between waveguides by coherently exciting the symmetric state $|s\rangle$ through waveguide~S using a continuous tone at the frequency $\omega_s$. With increasing drive strength, we observe a Mollow triplet obtained by driving the symmetric state towards saturation \cite{vanloo2013b}, which we use as a reference to calibrate subsequent PSD measurements from waveguide~S, $S_s(\omega)$ (see Sec.~IV of Supplementary material). Similarly, we calibrate the PSD from waveguide~A, $S_a(\omega)$. Driving only through waveguide S at an amplitude corresponding to a Rabi frequency $\Omega_s / 2\pi = 1.47$ MHz, we simultaneously measure both $S_s(\omega)$ and $S_a(\omega)$ as a function of the dephasing rate [Fig.~\ref{fig3} (a)]. In the absence of dephasing, no photons are detected through waveguide~A, while $S_s(\omega)$ displays a broad resonance fluorescence spectrum, indicative of inelastically scattered photons [bottom orange line in Fig. \ref{fig3} (a)]. The emitted power from the two modes into their respective waveguides is obtained by integrating the measured PSD. For increased dephasing rates, the total power re-emitted from the symmetric mode decreases monotonically [Fig.~\ref{fig3}(b)]. By contrast, the power detected in waveguide~A initially rises sharply to a pronounced  maximum before it starts to decrease. The initial rise results from the system overcoming its energy mismatch through noise-induced incoherent transitions between the symmetric and antisymmetric states. We again verify that it is the frequency components of the noise that bridge the energy gap $2g$ that enables excitation transport by reducing the bandwidth of the noise, for which no change in power transfer is observed [inset in Fig.~\ref{fig3}(b)]. For increasing dephasing rates, the power transfer is suppressed because of population localization, referred to as the quantum Zeno effect, a known feature of noise-assisted transport \cite{rebentrost2009a, potocnik2018c}. 

We explain the observed power transfer by a model based on the Lindblad master equation (see Sec.~V of the Supplementary material). In the presence of dephasing of qubit 1, the system exhibits longitudinal coupling through the $\sigma_z^{1}$ operator to the spectral environment represented by $S_{\Phi}(\omega)$ in the flux-line. Expressed in the symmetric-antisymmetric basis this coupling takes the form $\frac{1}{2}(\sigma_z^{s} + \sigma_z^{a}) + \sigma_s^{+}\sigma_a^{-}+\sigma_a^{+}\sigma_s^{-}$. The initial two terms represent collective dephasing of the $|s\rangle$, $|a\rangle$ subspace, influenced by the zero-frequency component $S_{\Phi}(0)$ of the noise. Conversely, the subsequent cross terms enable interaction between the two modes, harnessing the frequency components at $S_{\Phi}(2g)$ to connect the two states, thereby enabling power transfer. Based on this model and on our independently obtained estimates of the system parameters, we calculate heat flows in excellent agreement with the measured ones [dashed lines in Fig.~3(b)].

Having specified the setup of our device and its capabilities for energy transport, we demonstrate its operation as a quantum thermal machine. Classical microwave noise is synthesized at room temperature and admixed with quantum vacuum fluctuations from a resistive network of microwave attenuators at different temperature stages \cite{scigliuzzo2020, fink2010a}. This thermal radiation is characterized by a spectral profile of square white noise centered around $\omega_s$ and $\omega_a$ [Fig. \ref{fig1} (c)]. The baths will thereby acquire an average thermal population of $n_A$ and $n_S$, dependent on the amplitude of the noise. We accurately determine the thermal populations for a given noise  amplitude by measuring their PSD, calibrated as above.

\begin{figure}
    \includegraphics[width=9.0cm]{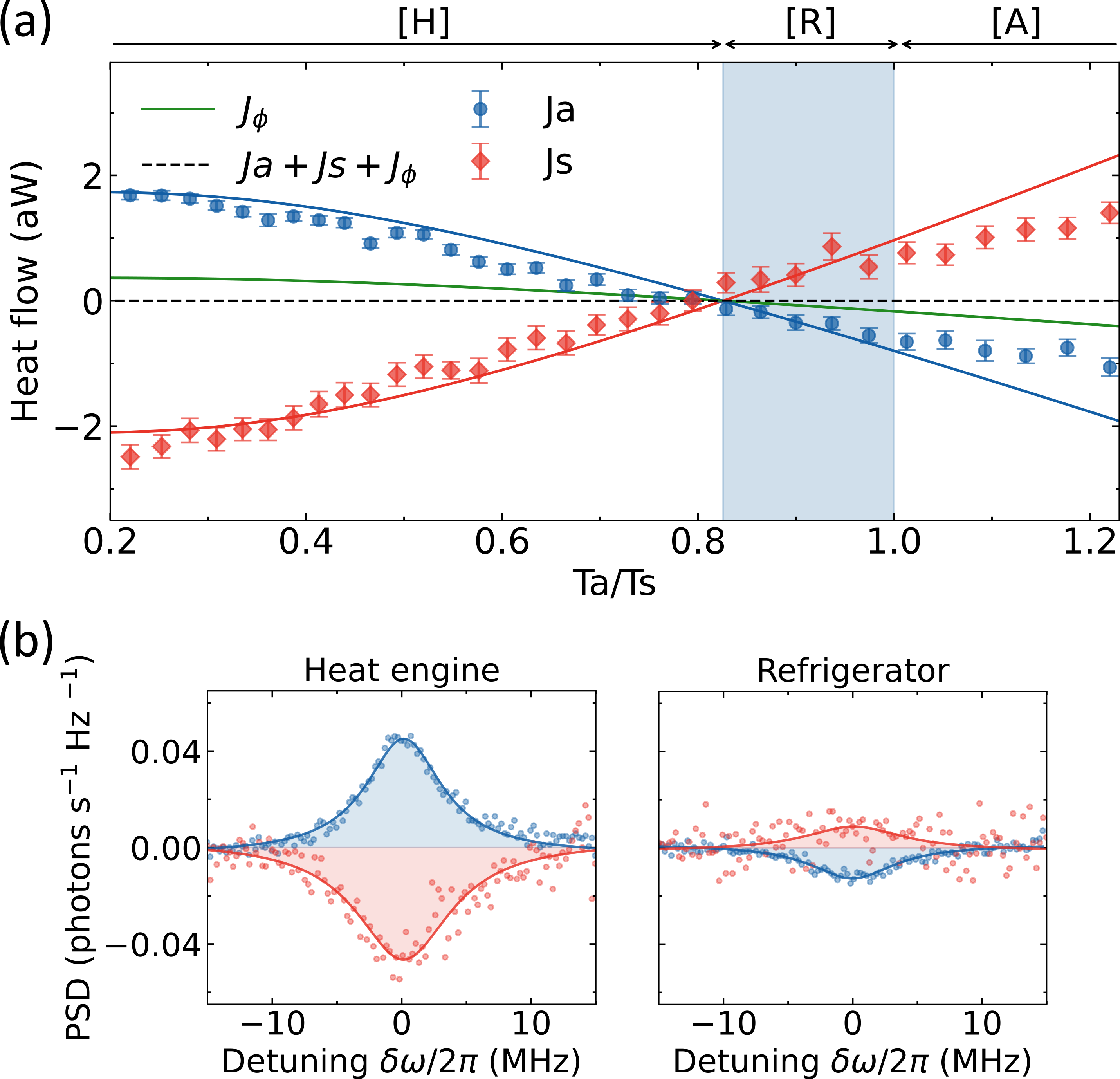}
    \caption{\label{fig4} Operational modes of our quantum thermal machine. (a) Heat flows through the antisymmetric (blue) and symmetric (red) waveguides as a function of the temperature ratio $T_a/T_s$ at a fixed dephasing rate $\Gamma_{\phi}$. Whilst $T_s$ remains fixed at 177 mK, $T_a$ is increased from base temperature to 217 mK. The first region [H] marks the operational regime of a heat engine, [R] that of a refrigerator (shaded blue region) and [A] a thermal accelerator. Solid lines are theoretical predictions from our analytical expression of the heat flows through the two waveguides and the dephasing channel (green solid line). The dashed line indicate the sum of the theoretical predictions $J_a + J_s + J_{\phi}$ (b) Differential power spectral density (PSD) measurement in the heat engine operational regime [H] at $T_a/T_s = 0.2$ and in the center of the refrigeration region [R] at $T_a/T_s = 0.92$.}
\end{figure}

We keep a constant dephasing rate $\Gamma_{\phi}/2\pi = 0.94$ MHz and the temperature of the symmetric waveguide at $T_s = 177$ mK.
We linearly increase the antisymmetric bath temperature from $T_a = 39$ mK (the minimum temperature we achieve in our waveguide \cite{aamir2023b}) to $T_a = 217$ mK, varying the temperature ratio $T_a/T_s$ in the range of 0.22 to 1.23, and measure the corresponding heat flows into each waveguide [Fig.~\ref{fig4} (a)].
For most temperatures, we measure heat flowing into the colder waveguide and out of the hotter one. However, our measurements indicate a temperature region
in which the heat flows are reversed compared to the temperature gradient, signalling refrigeration [shaded region in Fig.~\ref{fig4} (a)].

To elucidate the mechanism behind heat transport in our system, consider the interplay between separate thermalization to the baths and the effect of dephasing. Without dephasing, the symmetric (antisymmetric) transition thermalizes to waveguide S (A). The steady-state occupations of states $\ket{a}$ and $\ket{s}$ are, therefore,  $P_{\{a,s\}} \approx n_{\{a,s\}}$, where we have taking the limit $n_{\{a,s\}}\ll1$ for  simplicity of discussion. The dephasing noise act like an infinite-temperature bath for the $\{|s\rangle,|a\rangle\}$ subsystem, driving it  towards an equal mixture of $|s\rangle$ and $|a\rangle$. Depending on the initial populations before dephasing-induced mixing, the modes either absorb or emit photons to/from their respective baths to reestablish thermal equilibrium, evidenced by either a peak or a dip in their  PSD [Fig. \ref{fig4} (b)]. This interplay between separate thermalization and population balancing due to dephasing
manifests as heat flow. For the system to operate as a refrigerator, with the antisymmetric mode linked to the colder reservoir $T_a < T_s$, the initial population condition must satisfy $P_a > P_s$. Treating the modes as separate two-level systems, their average populations follow Fermi-Dirac statistics, leading to the refrigeration criterion $\omega_a/\omega_s < T_a/T_s < 1$. For our device, $\omega_a/\omega_s = 0.83$, which remarkably coincides, within experimental uncertainty, with the observed threshold value at which the sign of the heat currents is reversed and the device transitions into the refrigeration mode.

Numerical simulations based on Lindblad master equation quantitatively reproduce the observed heat flows [blue and red lines in Fig.~\ref{fig4} (a)]. Based on the theory model, we also calculate the heat flow through the dephasing channel $J_{\phi}$ (green line). Considering all the heat flows, three distinct operational regions emerge as the ratio $T_a / T_s$ increases. In the region where $0< T_a/T_s < 0.83$, the system operates as a heat engine [H] with $J_s < 0$ and $J_a, J_{\phi} > 0$ as defined in Reference \cite{buffoni2019}. In the interval $0.83< T_a/T_s < 1$ the antisymmetric bath temperature remains cooler than the symmetric bath and the heat flows are $J_a, J_{\phi} < 0$ and $J_s > 0$. In this configuration, the dephasing channel supplies the necessary work to drive heat flow against the temperature gradient and the system functions as a refrigerator [R]. As the temperature ration surpasses $T_a/T_s > 1$, the formerly colder bath becomes the hotter one, leading to heat flow from bath A (now hot) to bath S (now cold). In contrast to the heat engine, which exhibits a similar energy flows, energy input from the dephasing channel is required due to $\omega_a$ being smaller than $\omega_s$. Consequently, this regime aligns with the operational mode of a thermal accelerator [A], characterized by heat flows $J_a, J_{\phi} < 0$ and $J_s > 0$.

A key metric for assessing a refrigerator's performance is its coefficient of performance COP$ = \frac{J_a}{J_s-J_a}$, which gauges the efficiency of heat transfer from the cold to the hot reservoir \cite{chen2011b, hofer2016b}. We achieve a COP of 4.68, upper bounded by the Carnot limit COP$_{\text{Carnot}} = \frac{T_a}{T_s-T_a} = 4.88$ obtained from the temperatures at the crossover point between the heat engine and refrigerator. A significant factor influencing the efficiency of the refrigerator is the coupling rate $g$ between the transmons. Theoretical modelling shows a sharp decrease in COP going from smaller to larger coupling rates with a convergent behaviour towards COP$_{\text{Carnot}}$ for larger $g$ (see Sec.~V of Supplementary material).

In a proof of principle experiment, we studied an autonomous Brownian quantum refrigerator. We demonstrated how the interplay between individual thermalization and noise-assisted excitation transport can facilitate heat flow in steady state, both with and against a temperature gradient. Our thermal baths are composed of classically synthesized microwave noise with finite bandwidth and quantum vacuum fluctuations, enabling the customization of their spectral properties and selective heating of specific energy transitions. With precise control over device parameters, our machine can function as a heat engine, thermal accelerator, or refrigerator. While our method efficiently utilizes synthetic thermal baths, the setup is adaptable to natural thermal sources. Possible natural sources include the thermal radiation from various temperature stages of a dilution refrigerator \cite{wang2021k} or a heated resistor \cite{gubaydullin2022a}, preferably in combination with infrared-blocking filters that are transparent in the band of interest to prevent the generation of quasiparticles \cite{rehammar2023}.
Additionally, the investigation into systems similar to ours, designed for heat transport between thermal reservoirs have shown the theoretical possibility to violate the classical thermodynamical uncertainty relation (TUR) \cite{kalaee2021}. In the context of a quantum engines, this relation posits that fluctuations in the output power is inherently bounded by the entropy production of the system. The experimental validation of a classical TUR violation would mark the first demonstration of a quantum advantage in thermodynamics, although it partially depends on the ability to measure extremely small fluctuations in the heat currents between two thermal baths. Hence, our work paves a path towards further experimental studies within the field of quantum thermodynamics, notably in the realm of measuring heat flows between separated thermal reservoirs. Such measurements would be crucial, not only for investigating potential violations of the classical TUR but also in quantifying the energy costs of timekeeping \cite{erker2017a}.

\begin{acknowledgments}
The presented device design was assisted by the Python package QuCAT \cite{gely2020b} and was fabricated in Myfab Chalmers, a nanofabrication laboratory. This work received support from the Swedish Research Council via Grant No. 2021-05624, the Knut and Alice Wallenberg Foundation through the Wallenberg Center for Quantum Technology (WACQT), from the European Research Council via Grant No. 101041744 ESQuAT, and from the European Union via Grant No. 101080167 ASPECTS.
\end{acknowledgments}

%

\pagebreak
\widetext
\newpage
\begin{center}
\textbf{\large Supplemental Materials: Quantum refrigeration powered by noise in a superconducting circuit}
\end{center}
\setcounter{equation}{0}
\setcounter{figure}{0}
\setcounter{table}{0}
\setcounter{page}{1}
\makeatletter
\renewcommand{\theequation}{S\arabic{equation}}
\renewcommand{\thefigure}{S\arabic{figure}}
\renewcommand{\bibnumfmt}[1]{[S#1]}

\section{Full experimental setup}

\begin{figure}[h]
    \includegraphics[width=16.8cm]{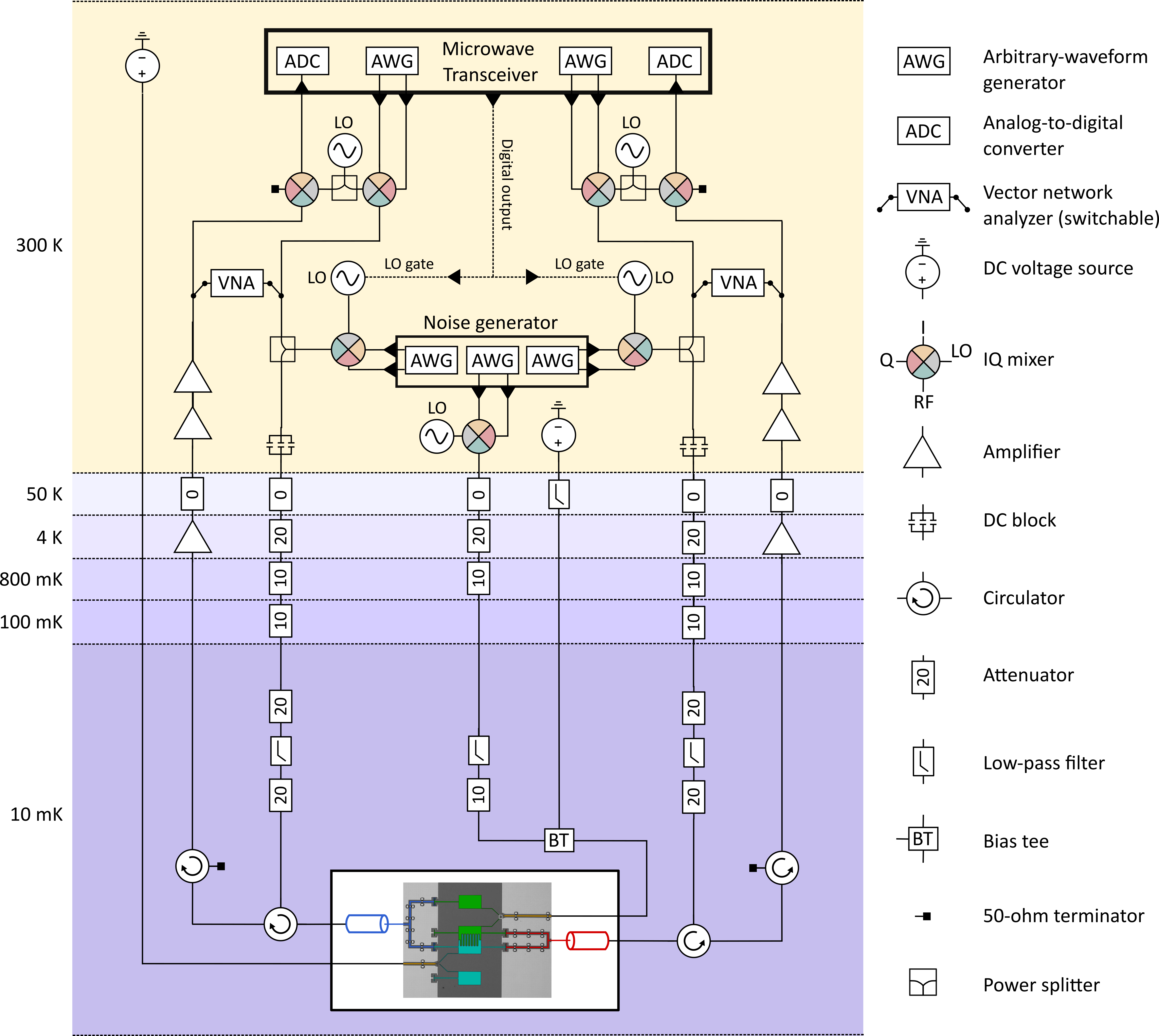}
    \caption{\label{expSetup} Experimental setup, see text for description.}
\end{figure}
The experimental setup for our study is depicted in Fig. \ref{expSetup}. Our device is placed at the mixing chamber stage of a dilution refrigerator, maintaining a stable temperature of 10 mK. To isolate from external interference's, the device is housed within a copper box which in turn is encased in a copper enclosure for electromagnetic wave shielding and a $\mu$-metal enclosure for protection against low-frequency magnetic fields.

The routing of input and output signals is managed by a microwave circulator, enabling a reflection measurement setup to both the symmetric and antisymmetric waveguide. For the input lines going to our device, and to some extent the RF-line combining with the flux line going to qubit 1, we employ highly attenuated coaxial lines. The signals emanating from the device are then captured in output lines equipped with a High Electron Mobility Transistor (HEMT) amplifier at the 4 K stage. Additional amplifiers are used at room temperature (300K) to further amplify the signal.

For our measurements we physically toggle between (indicated by a switch in Fig. \ref{expSetup}) a vector network analyzer (VNA) and a microwave transceiver where the latter is used in conjunction with in-phase-quadrature (IQ) mixers. The VNA is used for continuous-wave reflection spectroscopy and the microwave transceiver for frequency-resolved measurements. For the microwave transceiver we utilize a Quantum Machines OPX+ which is comprised of both arbitrary waveform generators (AWG) and analog-to-digital converters (ADC). Their functionality are enhanced with FPGA logic, facilitating both interleaved measurements and math operations such as the ability to compute the power spectral density of a time signal. 

To populate the waveguides with thermal noise we utilize arbitrary waveform generators of the Keysight 3202A model to continuously synthesize white voltage noise with a finite bandwidth of 30 MHz, a flat spectral density and centered at 200 MHz. The equipment is limited to a 500 MHz bandwidth so the noise is up-converted using IQ-mixers and local oscillators (Anapico APMS20G-4-ULN). The resulting continuous thermal radiation is segmented into pulses by gating the LO, coordinated by the digital outputs from the microwave transceiver playing simultaneously as the 5 $\mu$s readout events. This approach enables interleaved measurements of the power spectral density both with and without the applied noise. By doing so, the added noise can be directly subtracted from the background thermal radiation, thereby isolating the effects of our applied fields. When instead injecting coherent tones to the system, the AWG inside of the microwave transceiver are used. Similarly, the noise injected into the flux line going to qubit 1 is generated in the same way. It is combined with the DC current in the flux line using a cryogenic bias-tee.

\section{Reflection spectroscopy: model and supplementary data}

\begin{figure}[h]
    \includegraphics[width=12cm]{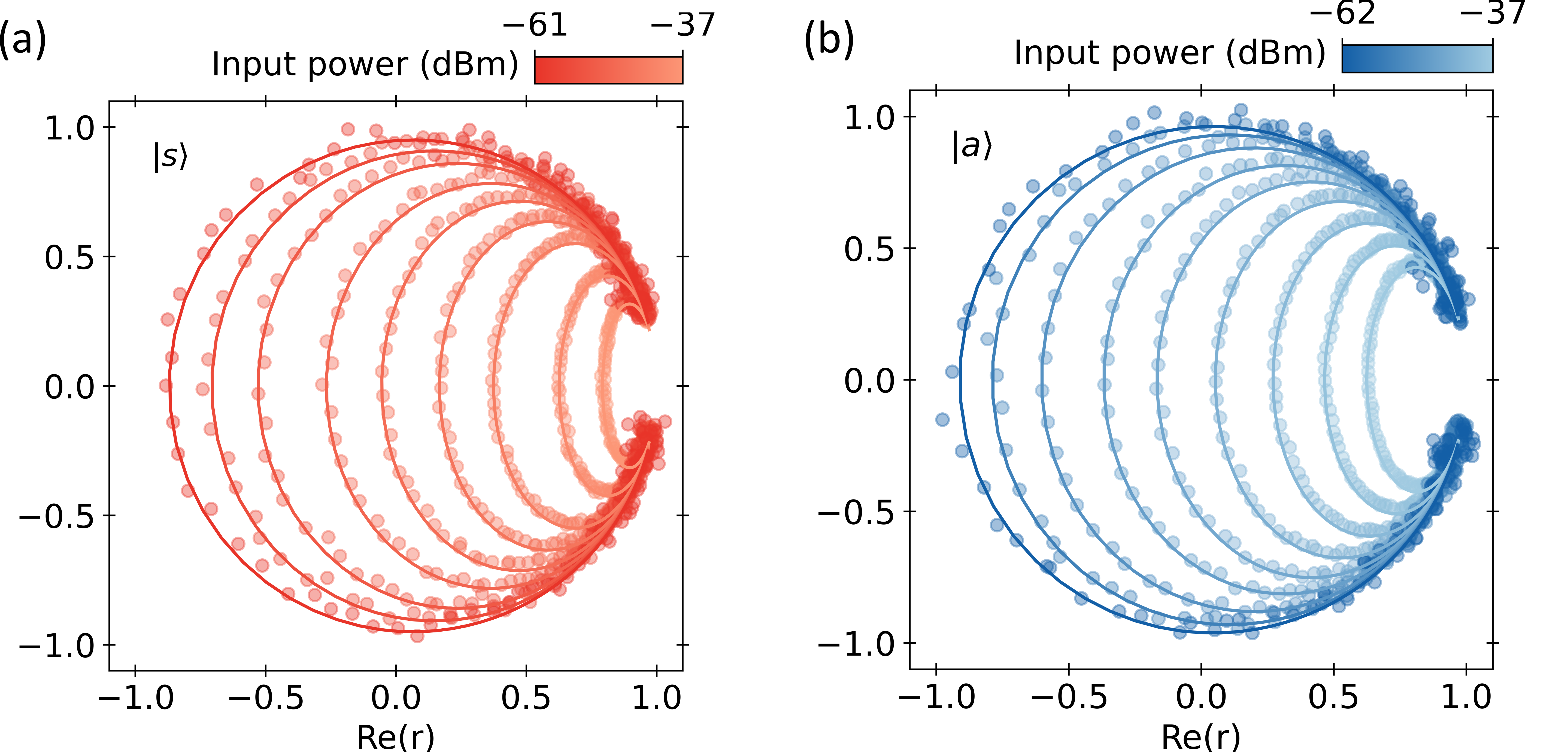}
    \caption{\label{reflection} Circular fit of real and imaginary part of power-dependent reflection coefficient, $r$, spectroscopy through (a) waveguide~S and (b) waveguide~A.}
\end{figure}
Using the Lindblad master equation and input-output theory, the reflection coefficient $r$  for a two-level system connected to the end of a waveguide can be derived \cite{Slu2021b, scigliuzzo2020}. For our device, the first states of the single excitation manifold, corresponding to symmetric and antisymmetric modes, are modeled as two-level systems. The reflection for each mode $i = \{s,a\}$ is described by the equation:
\begin{equation}
r\left(\omega-\omega_i\right)=1-\frac{i \Gamma_i \Gamma_{1 i}\left(\omega-\omega_i-i \Gamma_{2 i}\right)}{\Omega_i^2 \Gamma_{2 i}+\Gamma_{1 i}\left[\left(\omega-\omega_i\right)^2+\Gamma_{2 i}^2\right]}.
\end{equation}
In this expression, $\Gamma_i$ signifies the coupling rate between the state of symmetry $i$ and and the corresponding waveguide while $\Gamma_i^\prime$ is the coupling rate to all other decay channels fond in $\Gamma_{1 i}=\Gamma_i+\Gamma_i^{\prime}$ and $\Gamma_{2 i}=\left(\Gamma_i+\Gamma_i^{\prime}\right) / 2+\Gamma_{i \phi}$. Here $\Gamma_{i\phi}$ is the pure dephasing rate. Circular fits of the reflectance as a function of input power around the mode frequencies $\omega_{s,a}$ are shown in Figure \ref{reflection}. At low power the reflectance goes towards a near unit circle in the IQ plane while reducing towards a single point for lower powers. This trend signifies the shift from coherent to incoherent scattering observed as a two-level system approaches saturation \cite{scigliuzzo2020}.

\section{Reflection spectroscopy: Response to dephasing noise}
\begin{figure}[H]
\centering
    \includegraphics[width=16cm]{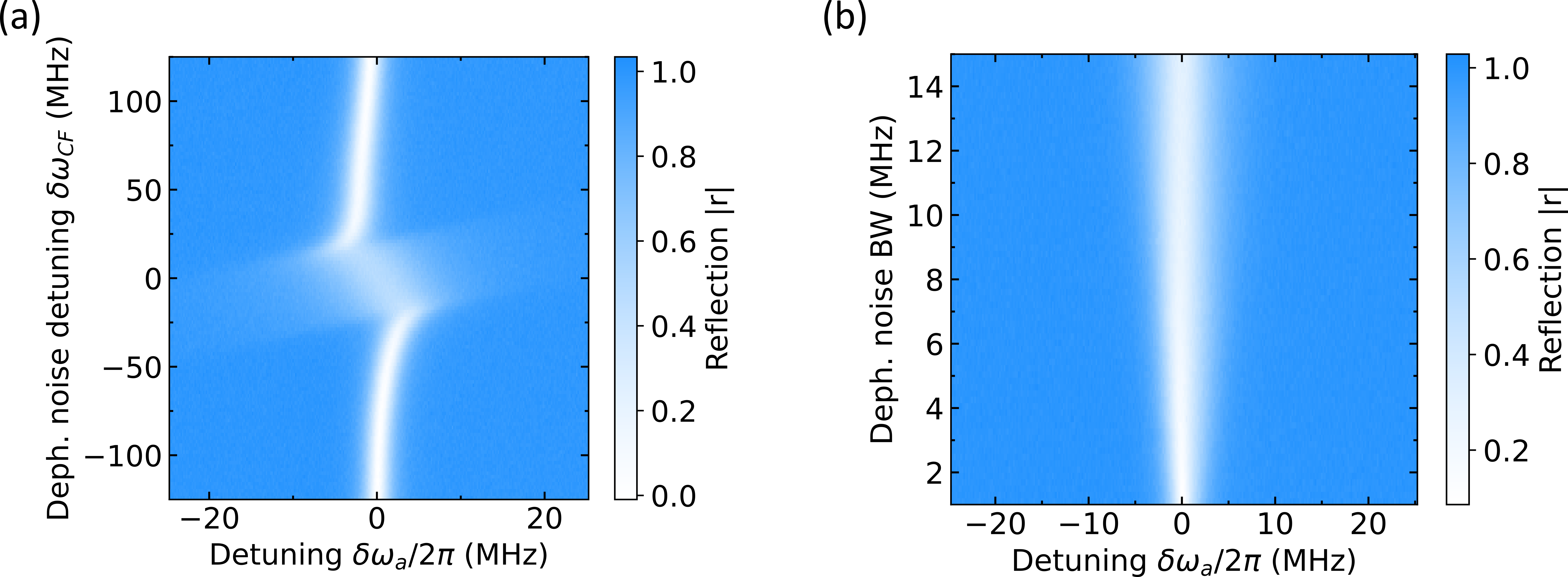}
    \caption{\label{SweepNoise1} Noise dependent reflection spectroscopy $|r|$ through the antisymmetric waveguide. (a) Reflection $|r|$ depending on the center frequency $\omega_{CF}$ of the applied noise for a fixed amplitude and a finite bandwidth of 50 MHz. (b) Reflection $|r|$ for increasing noise bandwidth for a fixed amplitude and $\omega_{CF} = 2g$.}
\end{figure}

The reflection in the antisymmetric waveguide is examined under varying noise spectrum characteristics. White noise with a square spectral profile, a constant amplitude equivalent to $\Gamma_{\phi}/2\pi = 1.28$ MHz and a finite bandwidth of 50 MHz is generated. The center frequency $\omega_{CF}$ of the noise is varied around the energy difference $2g$ [\ref{SweepNoise1}. (a)]. Notably, when the noise spectrum overlaps with the $2g$ frequency gap between the two modes, the linewidth of the $|0\rangle \rightarrow |a\rangle$ transition exhibits significant broadening. Furthermore, by increasing the noise bandwidth while maintaining a fixed amplitude (corresponding to $\Gamma_{\phi}/2\pi = 1.0$ MHz) and centering the frequency at $\omega_{CF} = 2g$ the linewidth saturates close after the bandwidth exceeds $\Gamma_A$ [\ref{SweepNoise1}. (b)].  

\section{Calibration of power measurements}
\begin{figure}[H]
\centering
    \includegraphics[width=14cm]{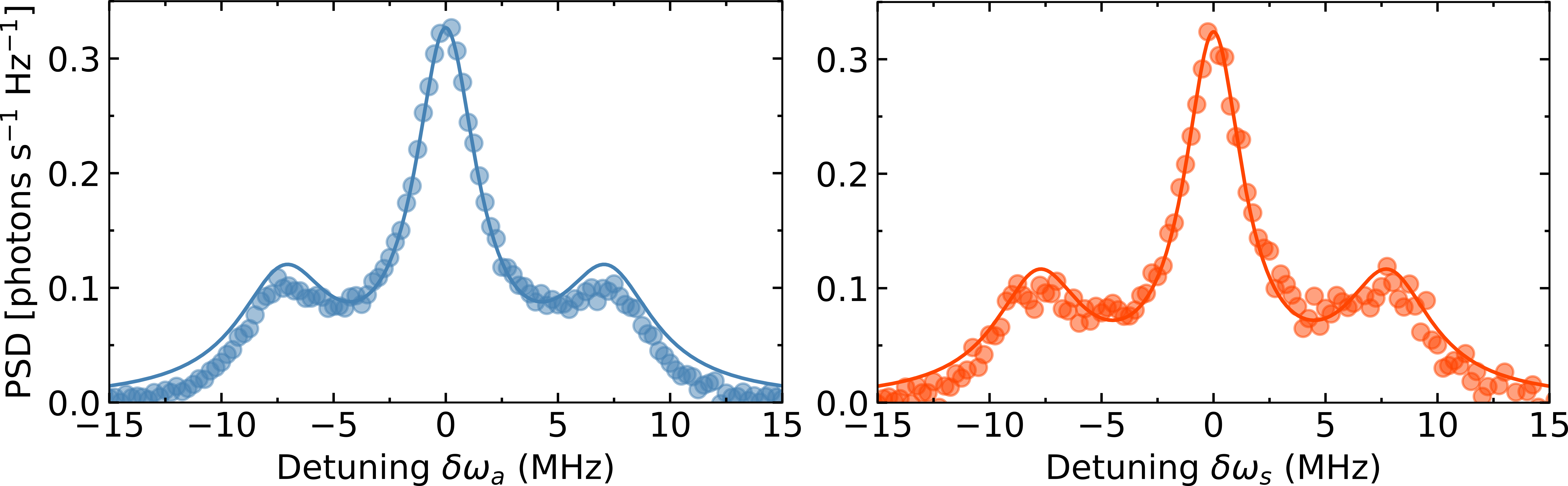}
    \caption{\label{MollowTriplet} Antisymmetric state- and symmetric state Mollow triplet used for calibration of the power spectral density measurements.}
\end{figure}

The thermal population resulting from the introduction of synthesized thermal fields was determined through direct measurements of their power spectral density (PSD). The PSD measured from both waveguides was calibrated to accurately represent the bosonic environment within each waveguide. This calibration was achieved by measuring the power spectrum $S_s(\omega)$ and $S_a(\omega)$ when coherently driving the symmetric and antisymmetric state respectively. At higher power levels of the applied microwave tone, a fully resolved Mollow triplet for both the symmetric and antisymmetric states becomes evident, featuring distinctly resolved side peaks [Fig. \ref{MollowTriplet}]. Using the previously obtained radiative and non radiative coupling rates while assuming negligable dephasing, we fit the the resonance fluorescence spectrum data to the Mollow triplet expression \cite{vanloo2013b}. This is done using only the Rabi frequency $\Omega$ and the y-axis scaling of the data as variable parameters. The scaling factor derived from this fitting process is then applied to adjust all subsequent PSD measurements. This adjustment ensures that the PSD accurately represents the spectral environment within the waveguides prior to any changes introduced by the amplification chain.

With the power spectral density calibrated, it becomes possible to directly determine the thermal population corresponding to a varying amplitude of the injected synthesized thermal baths by measuring their scaled PSD. An arbitrary amplitude can be mapped to a specific photon number by fitting a few data points to a simple exponential model of the form $n(\alpha) = B|\alpha|^2$, where $\alpha$ is the amplitude and B is a scaling factor.
The power is directly obtained as the integral of the recorded time trace squared from each waveguide, in accordance with Parseval's theorem $\int_{0}^{\infty}|f_{\{s,a\}}(t)|^2 d t = \int_{-\infty}^{\infty}|S_{\{s,a\}}(\omega)|^2 d \omega$. This is computationally inexpensive and is calibrated against the integral of a known PSD measurement such as the Mollow triplet.

\section{Theoretical model: Heat flows with baths}

\begin{table}[H]
\centering
\begin{tabular}{c|c|c|c}

\hline Eigenstate & Bare states composition & Eigenvalue & Value/2 $\pi$ \\

\hline$|0\rangle$ & $|0,0\rangle$ & 0 & $0 \mathrm{GHz}$ \\

$|a\rangle$ & $|1,0\rangle-|0,1\rangle$ & $\omega-g$ & $5.305 \mathrm{GHz}$ \\

$|s\rangle$ & $|1,0\rangle-|0,1\rangle$ & $\omega+g$ & $6.426 \mathrm{GHz}$ \\

$|2+\rangle_L$ & $|2,0\rangle+|0,2\rangle-\frac{\alpha+\sqrt{16 g^2+\alpha^2}}{2 \sqrt{2} g}|1,1\rangle$ & $\frac{1}{2}\left(4 \omega+\alpha-\sqrt{16 g^2+\alpha^2}\right)$ & $10.542 \mathrm{GHz}$ \\

$|2-\rangle$ & $|2,0\rangle-|0,2\rangle$ & $2 \omega+\alpha$ & $11.598 \mathrm{GHz}$ \\

$|2+\rangle_U$ & $|2,0\rangle+|0,2\rangle- \frac{\alpha-\sqrt{16 g^2+\alpha^2}}{2 \sqrt{2} g}|1,1\rangle$ & $\frac{1}{2}\left(4 \omega+\alpha+\sqrt{16 g^2+\alpha^2}\right)$ & $12.787 \mathrm{GHz}$ \\

\hline
\end{tabular}
\caption{\label{eigenstates} Eigenstates and eigenvalues of the diagonalized Hamiltonian. Analytical expressions found in reference \cite{aamir2022a}}
\end{table}

In the case of two ideally hybridized qubits where $\omega_1 = \omega_2 = \omega$, the Hamiltonian of the system can be well approximated by 

\begin{equation}
    \mathcal{H}= \sum_{i=1,2}\omega \sigma_i^{+} \sigma_i^{-}+g\left(\sigma_1^{+}\sigma_2^{-}+\sigma_2^{+}\sigma_1^{-}\right).
\end{equation}

Here $\omega$, $\sigma_i^{+}$ and $\sigma_i^{-}$ are the bare mode frequency, creation and annihilation operators of qubit $i = {1,2}$ respectively and g is the inter-transmon coupling rate. When adding the anharmonicity $\alpha$ of the transmons as an additional term to the Hamiltonian; $\sum_{i=1,2}\frac{\alpha}{2} \sigma_i^{+}\sigma_i^{+}\sigma_i^{-}\sigma_i^{-}$, the eigenstates and eigenvalues of the diagonalized Hamiltonian is found in the first- and second excitation manifold as presented in Table \ref{eigenstates}. From spectroscopy data we found $\omega/2\pi = 5.866$ GHz, $\alpha/2\pi = 133$ MHz and $g/2\pi = 560.1$ MHz when assuming identical transmons. 

In the rotating frame, the creation and annihilation operators linked to the symmetric (S) and antisymmetric (A) modes are given by $\sigma_{\mathrm{s}}^{ \pm}=\left(\sigma_1^{ \pm}+\sigma_2^{ \pm}\right) / \sqrt{2}$ and $\sigma_{\mathrm{a}}^{ \pm}=\left(\sigma_1^{ \pm}-\sigma_2^{ \pm}\right) / \sqrt{2}$. The two waveguides couple to the $|0\rangle \rightarrow |s\rangle$ and $|0\rangle \rightarrow |a\rangle$ transitions with a rate $\Gamma_s$ and $\Gamma_a$ respectively, populated with a photon number denoted by $n_s$ and $n_a$. 

In the presence of dephasing of qubit 1, apart from the transverse coupling to the thermal reservoirs, the system exhibits longitudinal coupling to the spectral environment represented by $S_{\phi}(\omega)$ in the flux-line. This is done through the $\sigma_z^{1} = \sigma_1^{+}\sigma_1^{-} - \sigma_1^{-}\sigma_1^{+}$ operator. Expressing the coupling in the symmetric- antisymmetric basis gives us
\begin{equation}
    \sigma_z^{1} = \frac{1}{2}(\sigma_z^{s} + \sigma_z^{a}) + \sigma_s^{+}\sigma_a^{-}+\sigma_a^{+}\sigma_s^{-}.
\end{equation}
Here the initial two terms represent pure dephasing where both the symmetric and antisymmetric modes "wobble" together with respect to the ground state. The rate at which this occur is proportional to the zero-frequency component $S_{\phi}(0)$ of the noise. In contrast, the following cross terms enable excitation exchange at a rate $\Gamma_{\phi}$ between the two modes, harnessing the frequency components at $S_{\phi}(2g)$ which bridges the energy gap between the two states. Because the noise injected into the flux line have a finite bandwidth of 50 MHz and centered at $2g$, the effect of the noise is well approximated by only taking the cross terms into account. The full unitary dynamics relevant for heat flow calculations between the two reservoirs is thereby described by the Lindblad master equation
\begin{equation}
    \dot{\rho}=\mathcal{L} \rho=-i[\mathcal{H}, \rho]+\mathcal{L}_s \rho+\mathcal{L}_a \rho+\frac{\Gamma_\phi}{2} \mathcal{D}\left[\sigma_s^{+}\sigma_a^{-}+\sigma_a^{+}\sigma_s^{-}\right] \rho.
    \label{ME}
\end{equation}

Here $\rho$ is the density matrix and  $\mathcal{D}[X]$ is the Lindblad super operator

\begin{equation}
   \mathcal{D}[X] \rho = X \rho X^{\dagger}-\frac{1}{2} X^{\dagger} X \rho-\frac{1}{2} \rho X^{\dagger} X.
\end{equation}

The terms $\mathcal{L}_s$ and $\mathcal{L}_a$ contains the dissipation taking place between the waveguides and their respective state transition

\begin{equation}
    \mathcal{L}_{j=\{s, a\}} \rho=\Gamma_j\left(n_j+1\right) \mathcal{D}\left[\sigma_j^{-}\right] \rho+\Gamma_j n_j \mathcal{D}\left[\sigma_j^{+}\right] \rho.
\end{equation}

In steady state, the change in energy through the two waveguides in addition to the longitudinally coupled dephasing channel, will vanish \cite{Sbalachandran2019a}

\begin{equation}
    \frac{d}{d t}\langle \mathcal{H}\rangle=\text{Tr}(\mathcal{H} \mathcal{L} \rho)=0.
\end{equation}
The individual contribution through the different channels for small thermal populations $n_s$ and $n_a$, can therefore be calculated as

\begin{equation}
\begin{split}
J_{i=\{s, a\}} &=\operatorname{Tr}\left(\mathcal{H} \mathcal{L}_i  \rho\right) \\
& \approx \frac{\hbar\left(n_a-n_s\right) \Gamma_a \Gamma_s \Gamma_\phi\left(g \pm \omega\right)}{\Gamma_s \Gamma_\phi+\Gamma_a\left(2 \Gamma_s+\Gamma_\phi\right)}
\end{split}
\end{equation}

where the expression has been reduced to a linear expression of the populations through a first order expansion in $n_s$ and $n_a$. Similarly;

\begin{equation}
\begin{split}
J_{\phi} &=\operatorname{Tr}\left(\mathcal{H} \frac{\Gamma_\phi}{2} \mathcal{D}\left[\sigma_s^{+}\sigma_a^{-}+\sigma_a^{+}\sigma_s^{-}\right] \rho\right) \\
&\approx  -\frac{2 g \hbar\left(n_a-n_s\right) \Gamma_a \Gamma_s \Gamma_\phi}{\Gamma_s \Gamma_\phi+\Gamma_a\left(2 \Gamma_s+\Gamma_\phi\right)}.
\end{split}
\end{equation}
For the simulations shown in Fig. 4a in the main text, we use experimentally determined system parameters to obtain numerical values of the heat flows.

\begin{figure}[h]
    \includegraphics[width=8.5cm]{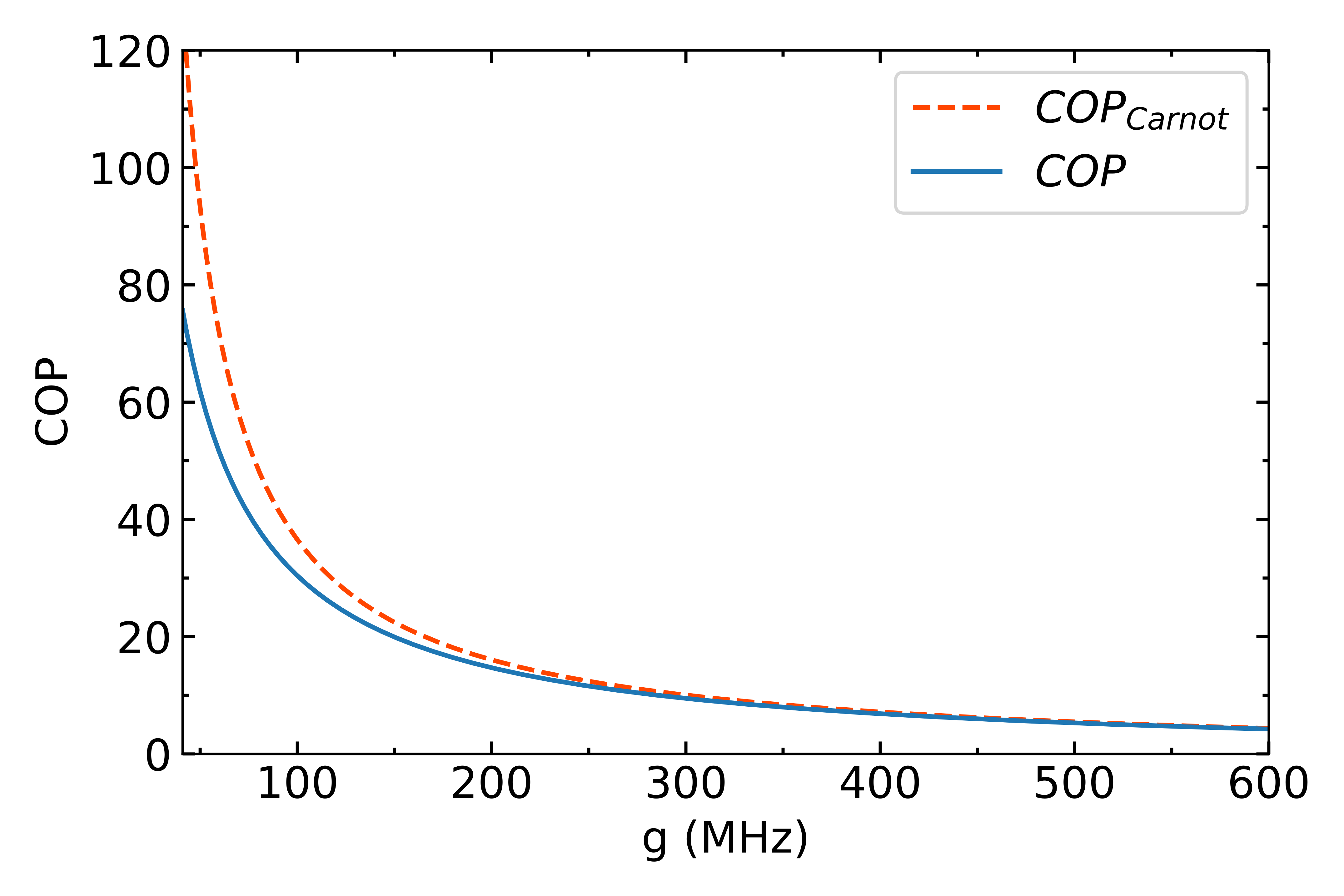}
    \caption{\label{Carnot} Coefficient of performance (COP) for the device (blue) in the refrigeration regime and the Carnot limit COP$_\text{Carnot}$ as indicated by the dashed orange line.}
\end{figure}

In the refrigeration region when the antisymmetric waveguide acts as the cold bath the coefficient of performance reads COP $ = \frac{J_a}{J_s - J_a}$ compared to the Carnot limit to COP$_{\text{Carnot}} = \frac{T_a}{T_s - T_a}$ \cite{chen2011b}. In the limit for large coupling rates and at the intersection between the heat engine and refrigeration regime, the COP for our device converges towards COP$_{\text{Carnot}}$  (Fig. \ref{Carnot}). 

\section{Numerical simulations of driven system}
In the presence of a coherent drive with a frequency $\omega_d = \omega_s$ going into the symmetric waveguide, the governing Hamiltonian can be expressed as:
\begin{equation}
    \hat{\mathcal{H}}= \sum_{i=1,2}\omega \sigma_i^{+} \sigma_i^{-}+g\left(\sigma_1^{+}\sigma_2^{-}+\sigma_2^{+}\sigma_1^{-}\right) + \frac{\Omega_s}{2} \left(\sigma_s^{+} + \sigma_s^{-} \right)
\end{equation}
The unitary dynamics are described by the Lindblad master equation in equation \ref{ME} but with the dissipators reflecting the absence of added thermal baths. Assuming negligible thermal population we have
\begin{equation}
    \mathcal{L}_{j=\{s, a\}} \rho=\Gamma_j \mathcal{D}\left[\sigma_j^{-}\right] \rho.
\end{equation}

From the aforementioned definition of heat flow $J_a =\operatorname{Tr}\left(H \mathcal{L}_a  \rho\right)$, the power transfer into the antisymmetric waveguide can be calculated as a function of dephasing. This is done using the python library QuTip 4.7.1 \cite{Sjohansson2013b} through which the steady state density matrix can be calculated numerically. However, for the response in the symmetric waveguide, the power spectrum is calculated for the steady state solution when $\dot{\rho}=0$ from a two-time correlation function

\begin{equation}
S(\omega)=\int_{-\infty}^{\infty}\langle A(\tau) B(0)\rangle e^{-i \omega \tau} d \tau.
\end{equation}
For the emission through the symmetric waveguide we use the correlation function $\langle \sigma_s^{+}(\tau) \sigma_s^{-}(0)\rangle$. By multiplying with the radiative decay rate $\Gamma_s$ the correct magnitude is obtained, resulting in the power spectrum
\begin{equation}
S_s(\omega)=\Gamma_s \int_{-\infty}^{\infty}\langle \sigma_s^{+}(\tau) \sigma_s^{-}(0)\rangle e^{-i \omega \tau} d \tau.
\end{equation}
Using QuTip, this quantity is directly calculated and the emitted power is obtained by integrating $S_s(\omega)$ over a frequency span of 20 MHz. This is the same frequency span used in the measurements from the main text.

\end{document}